\documentstyle{article}

\font\tenrm=cmr10

\font\elevenbf=cmbx10 scaled\magstep 1
\font\elevenrm=cmr10 scaled\magstep 1
\font\elevenit=cmti10 scaled\magstep 1

\newcommand{\beq}{\begin{equation}}
\newcommand{\beqn}{\begin{eqnarray}}
\newcommand{\eeq}{\end{equation}}
\newcommand{\eeqn}{\end{eqnarray}}
\def\slash#1{\setbox0=\hbox{$#1$}#1\hskip-\wd0\hbox to\wd0{\hss\sl/\/\hss}}
\renewenvironment{thebibliography}[1]
 { \elevenrm
   \begin{list}{\arabic{enumi}.}
    {\usecounter{enumi} \setlength{\parsep}{0pt}
     \setlength{\itemsep}{3pt} \settowidth{\labelwidth}{#1.}
     \sloppy
    }}{\end{list}}

\begin{document}
\begin{center}{{\elevenbf $q\bar q$ BOUND STATES IN THE BETHE-SALPETER
 FORMALISM\\}
\vglue 1.0cm
{\elevenrm Pankaj Jain and Herman J. Munczek\\}
\vglue 0.5cm
\baselineskip=13pt
{\elevenit  Department of Physics and Astronomy \\}
\baselineskip=12pt
{\elevenit The University of Kansas\\}
\baselineskip=12pt
{\elevenit Lawrence, KS 66045-2151\\}
\vglue 2.0cm
{\elevenrm ABSTRACT}}
\end{center}
\vglue 0.3cm
{\rightskip=3pc
 \leftskip=3pc
 \tenrm\baselineskip=12pt
 \noindent
We solve the Bethe-Salpeter equation in order to determine the
spectrum of pseudoscalar and vector meson bound states
for light as well as heavy quarks. The fermion propagators are
obtained by solving the Schwinger-Dyson equation consistently with
the Bethe-Salpeter equation, a procedure necessary for demonstrating the
Goldstone nature of the pion in the chiral limit. Our results agree
qualitatively and quantitatively with expectations both from current
algebra for light quarks and from composite models for heavy quarks.}
\vfill
\noindent
Presented at the DPF92 meeting, Fermilab, Chicago, Nov. 10-14, 1992
\eject
{\elevenbf\noindent 1. Introduction}
\vglue 0.1cm
\baselineskip=14pt
\elevenrm
There has been considerable theoretical study of the strongly interacting
bound states. However most of the work is based on non-relativistic
approximations to the bound state
 equations. This approach should be accurate if
we consider the bound states which only involve heavy quarks, but
is expected to fail for light mesons. In particular it does not
take into account the fact that the light pseudoscalar mesons are
almost Goldstone particles and that the light quark masses are
mostly dynamical. Furthermore in most of the studies
there is no clear connection between the constituent quark mass
which is fed into the bound state equation and the current algebra
mass. In order to correctly describe the light mesons it is necessary
to go to a field theoretic framework.
In the present paper
we study the quark-antiquark bound states by solving the Bethe-Salpeter (BS)
 equation consistently with the Schwinger-Dyson (SD) equation
for the quark propagator, in such a way that we get
zero mass pseudoscalar states in the limit of zero bare quark masses.
 To simplify the problem we work in the
  ladder approximation. We employ for our calculations theoretically
and phenomenologically motivated models for the gluon propagator.

The BS equation in the ladder approximation can be written as,
\beq
S_a^{-1}(q+\xi p)\chi(p,q)S_b^{-1}(q-(1-\xi)p) = -i\int\gamma_\mu
\chi(p,k)\gamma_\nu G_{\mu\nu}(k-q){d^4k\over (2\pi)^4}\; ,
\eeq
The quark propagators are determined by solving the SD equation,
also in the ladder approximation,
\beq
 S_a^{-1}(q) = \slash q - \tilde m_a(\Lambda_c) + i\int\gamma_\mu S_a(k)
\gamma_\nu G_{\mu\nu}(k-q){d^4k\over (2\pi)^4}\; ,
\eeq
where $\Lambda_c$ is the ultraviolet cutoff.
The gluon propagator, $G_{\mu\nu}$, is modelled in terms of
several parameters which can be fitted to experimental data. It is
given in the Landau gauge
 by $G_{\mu\nu}(k)=-(g_{\mu\nu}-k_\mu k_\nu/k^2)G(k^2)$ where,
$$G(k^2) = {16\pi^2\over 3}\bigg\lbrack
{d\over k^2ln(x_0+x)}\biggl(1+b{ln[ln(x_0+x)]\over
ln(x_0+x)}\biggr) + (2\pi\eta)^2\delta^{(4)}(k)
+ a(1-\omega k^2/k_0^2)e^{-k^2/
k_0^2}\bigg\rbrack\; ,$$
and $x=k^2/\Lambda^2_{QCD}$. This model
goes to the two loop form of the running coupling at large momentum
and leads to an approximately harmonic
oscillator potential in three dimensional configuration space, a form which
is necessary to get the realistic spectrum of heavy mesons.
The BS
wave function for the pseudoscalar bound state can be expressed as,
\beq
\chi(p,q) = \gamma_5[\chi_{0} + \slash p\chi_{1} + \slash q\chi_{2}
+ [\slash p,\slash q]\chi_{3}]
\eeq
with similar decomposition for other spin and parity states. The
resulting BS equation is simplified by expanding the wave functions,
$\chi_i$ in terms of Tschebyshev polynomials,
\beq
\chi_i\biggl(q^2,M_B^2,\cos
\theta\biggr)=\sum\limits_n\chi_i^{(n)}\biggl(q^2,M_B^2\biggr)T^{(n)}
(\cos\theta)\;  ,
\eeq
where we have set $p^2=-M^2_B$, the bound state mass squared.
We keep only the leading order polynomial for our calculation.
The contribution due to the dropped terms was estimated to be small.
Details of the calculation for pseudoscalars are given in Ref. [1].
\vglue 0.6cm
{\elevenbf\noindent 2. Results and discussion}
\vglue 0.4cm
 Here we describe some qualitative and quantitative features of
the results for pseudoscalar and vector mesons.
The detailed numerical results for pseudoscalar mesons
and the parameter choices are given in Ref. [1].
The mass functions for different quarks are displayed
in Fig. 1. The asymptotic behavior of
the mass functions and the wave functions agrees with the one found on the
basis of operator-product-expansion considerations. For light mesons
our results satisfy the Gell-Mann, Oakes and Renner relation,
$M^2_{ab} = [\tilde m_a(\Lambda_c)+\tilde m_b(\Lambda_c)]<q\bar q>_{ch}/
f^2_{ch}$,
where $\tilde m$ is the bare mass of the quark defined in Eqn. 2 and
the subscript $ch$ means that the quantity is computed by using
$\tilde m(\Lambda_c)=0$. Both $\tilde m$ and the condensate $<q\bar q>_{ch}$
depends on the ultraviolet cutoff $\Lambda_c$ but their product was
found to be insensitive to $\Lambda_c$. For heavy quarks, the
relationship between $M_{ab}$ and $[m_a(0)+m_b(0)]$ is roughly linear, in
agreement with nonrelativistic limit expectations.

We have also obtained preliminary results for vector mesons ground states.
The overall fit was found to be better with
parameters choices, $a$=-[32.9 MeV]$^{-2}$,
$\eta$=270 MeV with the remaining parameters
same as in Ref. 1. The results for pseudoscalar mesons were found to be
within a few \% of the ones given in [1].
The results for some of the vector mesons are as follows:
$m_\rho$ = 787 MeV (770 MeV), $m_{K^*}$ = 978 MeV (892 MeV),
 $m_\phi$ = 1250 MeV (1019 MeV), $m_{D^*}$
= 2075 MeV (2010 MeV), $m_{B^*}$ = 5325 MeV (5324 MeV), $m_{J/\psi}$ =
3245 MeV (3097 MeV), $m_\Upsilon$ = 9500 MeV (9460 MeV).
 The numbers in parenthesis are the experimental
results. These results were obtained by keeping only the
dominant term in the invariant decomposition of the vector meson wave
function. More complete results will be given in a forthcoming
publication.

Finally we display our preliminary results for the electromagnetic and isospin
splittings for pseudoscalar and vector mesons. Our
results for the pseudoscalar mass splittings decrease significantly as we
go from the Kaon to the B meson, in reasonable agreement with experiments.
Quantitatively we find, $K^0-K^+$ = 4.02 MeV (input), $D^+-D^0$ = 3.95 MeV
(4.77 MeV)
$B^0-B^+$ = 1.0 MeV (0.1 MeV).
For the vector meson
we do not get as strong a decrease as is displayed by the
experiments. The results are as follows:
 $K^{*0}-K^{*+}$ = 11 MeV (6.7 MeV), $D^{*+}-D^{*0}$ = 10.8 MeV (2.9 MeV)
$B^{*0}-B^{*+}$ = 7.4 MeV (?)
 We are currently examining different models and
calculating corrections to our result to determine if the agreement can
be improved.

In conclusion, we have presented a covariant treatment of $q\bar q$ bound
states which is applicable for both light and heavy mesons.
Qualitatively our results
are in good agreement with current algebra results for
light quarks and nonrelativistic limit expectations for heavy quarks.
Further tests of the approach described here, as well as parametrization
of $\alpha_s(q^2)$, will be done by obtaining a more complete spectrum and
by computing the electromagnetic and
weak form factors of these mesons.
\vglue 0.6cm
{\elevenbf \noindent 5. Acknowledgements \hfil}
\vglue 0.4cm
We thank Douglas W. McKay and John P. Ralston for useful discussions.
This work was
supported in part by the Department of Energy under grant No.
DE-FG02-85-ER40214.
\vglue 0.5cm
{\elevenbf\noindent 6. References \hfil}
\vglue 0.4cm

\end{document}